# Quantisation of Deformed Special Relativity

CLEMENS HEUSON
*Zugspitzstr. 4*
*D-87493 Lauben, Germany*

Abstract:

We consider deformed special relativity (DSR) theories on commutative space-time, perhaps as an first approximation to a noncommutative space-time formulation. The corresponding field theories in general possess derivatives of all orders. From a quantisation procedure and the deformed Dirac equation in momentum space we obtain for a specific DSR model enhanced expressions for momentum and charge and finite vacuum contributions.

## 1. Introduction

Deformed or doubly special relativity theories, involving two invariant scales, a velocity scale and an fundamental length scale $\ell$ or energy scale $\kappa = 1/\ell$ (presumably related to the Planck length or mass), have attracted a lot of interest in the last years, see the review in [1]. There are also severe problems in the interpretation of DSR theories, the soccer ball problem (macroscopic bodies have energies exceeding the Planck energy), the question of the proper formulation in space-time (noncommutative or not), the multiplicity problem (which if any model describes nature), the issue of particle velocity etc., see for example the discussion in [2] and papers cited therein.

In this letter we consider second quantisation of DSR theories as a first step for a DSR quantum field theory. The plan of the paper is as follows. In section 2 we set up the formulation and notation of DSR theories. In section 3 the field theory on commutative space-time is considered, perhaps as a first approximation in the case of noncommuting space-time. In section 4 we discuss the deformed Dirac equation and plane wave solutions for an generic DSR theory. In section 5 the Lagrangian, equations of motion and expressions for momentum and charge are derived for a specific model. This serves in the next section to formulate a quantisation procedure, yielding deformed terms for momentum and charge and finite vacuum contributions to energy and charge. In the final section possible extensions are discussed.

## 2. Deformed Special Relativity theories

The starting point of any DSR theory is a deformed dispersion relation written in the form

$$F^2 E^2 - G^2 \boldsymbol{p}^2 = m^2 \tag{1}$$

with functions $F, G = f(E, \boldsymbol{p}^2, \kappa)$ preserving rotational symmetry. The mass $m$ on the right side of (1) is related to the rest mass $m_0$ by $m = m_0 F(m_0, 0, \kappa)$ and thereby the functions $F, G$ could always be redefined in a way that $m^2$ on the right side equals the rest mass $m_0^2$. Since the DSR symmetry group is considered to be a nonlinear realisation of the Lorentz group, we can find functions mapping the physical momentum $p = (E, \boldsymbol{p})$ into an unphysical momentum $\pi = (\varepsilon, \boldsymbol{\pi})$ transforming linearly under Lorentz transformations [3], [4].

$$\pi = U(p) \text{ or } \varepsilon = FE, \ \boldsymbol{\pi} = G\boldsymbol{p} \tag{2}$$

The inverse relations with the functions $\overline{F}, \overline{G} = g(\varepsilon, \boldsymbol{\pi}^2, \kappa)$ are given by

$$p = \overline{U}(\pi) \text{ or } E = \overline{F}\varepsilon, \ \boldsymbol{p} = \overline{G}\boldsymbol{\pi} \tag{3}$$

We used the slightly modified notation of [4], [5]. Since standard Lorentz transformations act linearly on the unphysical momentum $\pi' = \Lambda\pi$, the action of deformed Lorentz transformations on $p$ is given by

$$p' = \overline{U} \circ \Lambda \circ U(p) \tag{4}$$

The deformed infinitesimal generators $M_{\mu\nu}$ can be obtained from the undeformed generators $L_{\mu\nu}$, see [4]

$$M_{\mu\nu} = \overline{U} L_{\mu\nu} U, \ L_{\mu\nu} = i(p_\mu \partial_\nu - p_\nu \partial_\mu) \tag{5}$$

or simply by transforming $\widetilde{M}_{\mu\nu} = i(\pi_\nu \frac{\partial}{\partial \pi_\mu} - \pi_\mu \frac{\partial}{\partial \pi_\nu})$ to physical momenta [1]. Due to the assumed rotational symmetry only the boost generators $M_{0i}$ are modified.

The various DSR models considered so far, are specified by different functions $F, G$ in (1) with deformed Lorentz transformations constructed according to (4), see [1]-[6]. Examples for the functions $F, G$ in the models [7]-[10] can be found in the following table:

| DSR model | function F | function G |
|---|---|---|
| $\kappa$-Poincare [7] | $F^2 = 2\kappa^2 \cosh(E/\kappa)/E^2$ | $G^2 = \exp(E/\kappa)$ |
| Magueijo-Smolin [8] | $F = (1 - E/\kappa)^{-1}$ | $G = (1 - E/\kappa)^{-1}$ |
| Herranz [9] | $F = \kappa(\exp(E/\kappa) - 1)/E$ | $G = 1$ |
| Heuson [10] | $F = (1 - \boldsymbol{p}^2/\kappa^2)^{-1/2}$ | $G = (1 - \boldsymbol{p}^2/\kappa^2)^{-1/2}$ |

Table 1

A major problem of DSR theories remains proper the formulation in space-time. Motivated by the $\kappa$-Poincare model a space-time of the Lie algebra type was considered in [5]-[7], see also [11], [12]. There are indeed strong indications not only from DSR, that space-time becomes noncommutative at very small distances [13]. Then one has to refer to noncommutative field theory see [14] and references therein. The opposite view that space-time remains commutative has also been advocated, see for example [15], [16]. Here we consider a commutative space-time, which should at least serve as a first approximation for not to high energies and try to develop a field theoretical description as in [16].

## 3. Field Theory formulation of DSR theories

As discussed we assume a standard commutative space-time as a first approximation. The deformed dispersion relation (1) can then be translated into coordinate space by the replacement

$$p_\mu \to i\partial_\mu = i\frac{\partial}{\partial x^\mu} \tag{6}$$

Multiplying (1) by $\phi(x)$ one obtains the deformed Klein Gordon equation. Here we use the metric $\eta = diag(+,-,-,-)$ employed in most field theory books. But now as discussed in [16] a problem occurs: the dispersion relation (1) can be written in several ways, and these belong to different field theories. For example the Magueijo-Smolin model [7] with the dispersion relation $\frac{E^2 - \boldsymbol{p}^2}{(1-E/\kappa)^2} = m^2$ can be arranged as $E^2(1-m^2/\kappa^2) + (2m^2/\kappa)E - \boldsymbol{p}^2 = m^2$. By completing the square for $E$, one sees that this expression gives an energy independent redefinition of energy and mass and thereby a Klein-Gordon equation with highest second order derivatives. The original dispersion relation however corresponds to an equation of motion with infinitely many derivatives, as can be seen by substituting $p_\mu \to i\partial_\mu$ and Taylor expanding the denominator. A similar comment holds in the case of model [10]. The field theory models proposed in section IV of [16] originally yield a field theory with derivatives up to infinite order, while the rewritten form contains only derivatives up to fourth order. To avoid this ambiguity, we postulate that the dispersion relation must be written in the form $C = m^2$, where C is the left side of (1), i.e. the Casimir operator of the deformed Lorentz transformations in momentum space without any $m$. The equation of motion from (1) together with (6) is in general one with derivatives up to infinite order, of course also in the models [7], [9], excepted are cases where $F$, $G$ are polynomials in $p$. Thereby one encounters the difficulty of theories with infinitely many derivatives or approximately, by Taylor expanding and keeping some first terms, at least theories with derivatives of finite order greater than two. The field theory formulation of theories with higher derivatives is well known in the literature see for example [17] or [16]. For a Lagrangian $L$ with higher order derivatives with maximal order $N \le \infty$ the action is $S = \int d^4x L(\phi, \partial_\mu \phi, .., \partial_{\mu_1..\mu_N}\phi)$. The equations of motion are derived from $\delta S / \delta \phi = 0$:

$$\sum_{l=0}^{N}(-1)^l \partial_{\mu_1..\mu_l} \frac{\partial L}{\partial(\partial_{\mu_1..\mu_l}\phi)} = 0 \tag{7a}$$

$$\frac{\partial L}{\partial \phi} - \partial_{\mu_1} \frac{\partial L}{\partial(\partial_{\mu_1}\phi)} + ... + (-1)^N \partial_{\mu_1..\mu_N} \frac{\partial L}{\partial(\partial_{\mu_1..\mu_N}\phi)} = 0 \qquad (7b)$$

The invariance of the action $S$ under infinitesimal transformations of coordinates $x^\mu \to x^\mu + \delta x^\mu$ and fields $\phi \to \phi + \delta\phi$ gives Noether's conservation law $\partial_\mu j^\mu = 0$ with

$$j^\mu = \sum_{r=1}^{l} \delta\partial_{\mu_1..\mu_{r-1}}\phi \sum_{l=r}^{N} (-1)^{l-r} \partial_{\mu_r..\mu_{l-1}} \frac{\partial L}{\partial(\partial_{\mu_1..\mu_{l-1}\mu}\phi)} + L\delta x^\mu \qquad (8)$$

The pure Noether charge is obtained from (8) for $\delta x^\mu = 0$. The conserved energy momentum tensor is obtained for $\delta x^\mu = \delta\varepsilon^\mu$ and $\delta\phi = -\delta\varepsilon^\mu \partial_\mu \phi$ as

$$T^{\mu\nu} = \sum_{r=1}^{l} \partial^\nu_{\mu_1..\mu_{r-1}}\phi \sum_{l=r}^{N} (-1)^{l-r} \partial_{\mu_r..\mu_{l-1}} \frac{\partial L}{\partial(\partial_{\mu_1..\mu_{l-1}\mu}\phi)} - L\eta^{\mu\nu} \qquad (9a)$$

$$T^{\mu\nu} = (\frac{\partial L}{\partial(\partial_\mu\phi)} - \partial_{\mu_1} \frac{\partial L}{\partial(\partial_{\mu_1\mu}\phi)} + ..)\partial^\nu \phi$$
$$+ (\frac{\partial L}{\partial(\partial_{\mu_1\mu}\phi)} - \partial_{\mu_2} \frac{\partial L}{\partial(\partial_{\mu_1\mu_2\mu}\phi)} + ..)\partial^\nu_{\mu_1}\phi \qquad (9b)$$
$$+ ... - L\eta^{\mu\nu}$$

## 4. Deformed Dirac equation and free particle solutions

In this section we discuss the deformed Dirac equation and plane wave solutions used in the quantisation procedure. The deformed Dirac equation was derived in [18] for the $\kappa$-Poincare model. Here we consider it in commutative space-time for an generic DSR theory defined by dispersion relation (1) following [18] or [19]. Starting point is the observation that the Lorentz algebra of rotations $J_i$ and boosts $K_i$ is unchanged in DSR. This allows to introduce the operators $A_i = \frac{1}{2}(J_i + iK_i)$ and $B_i = \frac{1}{2}(J_i - iK_i)$ obeying two independent SU(2) algebra's. Left handed spinors $\phi_L$ with $A_i = 0$ and right handed spinors $\phi_R$ with $B_i = 0$ transform under a pure boost with rapidity $\xi$ as ($\sigma^i$ are the Pauli matrices)

$$\phi'_R = \exp(\sigma^i \xi_i /2)\phi_R \,,\, \phi'_L = \exp(-\sigma^i \xi_i /2)\phi_L \qquad (10)$$

Boosting from a rest frame one obtains

$$\phi_R(\boldsymbol{p}) = \exp(\sigma^i \xi_i /2)\phi_R(0) \,,\, \phi_L(\boldsymbol{p}) = \exp(-\sigma^i \xi_i /2)\phi_L(0) \qquad (11)$$

and with $\phi_L(0) = \phi_R(0)$ the relations

$$\phi_R(\boldsymbol{p}) = \exp(\sigma^i \xi_i)\phi_L(\boldsymbol{p}) \,,\, \phi_L(\boldsymbol{p}) = \exp(-\sigma^i \xi_i)\phi_R(\boldsymbol{p}) \qquad (12)$$

We use $\exp(\sigma^i \xi_i) = \cosh(\xi) + \sigma^i n_i \sinh(\xi)$, $n_i = p_i/|\boldsymbol{p}|$ and the deformed relations

$$\cosh(\xi) = FE/m \ , \ \sinh(\xi) = G|\boldsymbol{p}|/m \tag{13}$$

with $\psi = \begin{pmatrix} \phi_R \\ \phi_L \end{pmatrix}$ to obtain the DSR Dirac equation in momentum space from (12) as

$$(\gamma^0 FE + \gamma^i Gp_i - m)\psi(p) = 0 \tag{14}$$

where $\gamma^\mu$ are the Dirac matrices in Weyl representation [20] $\gamma^0 = \begin{pmatrix} 0 & 1 \\ 1 & 0 \end{pmatrix}$, $\gamma^i = \begin{pmatrix} 0 & \sigma^i \\ -\sigma^i & 0 \end{pmatrix}$.

The free particles solutions of (14) are constructed, assuming that the functions $F$, $G$ are rewritten so that $m$ coincides with the rest mass, see the comment following (1). The boost of a Dirac spinor $\psi$ from a rest frame is according to (11)

$$\psi' = S\psi \ , \ S = \begin{pmatrix} \exp(\sigma^i \xi_i / 2) & 0 \\ 0 & \exp(-\sigma^i \xi_i / 2) \end{pmatrix} \tag{15}$$

Now we insert $\exp(\pm \sigma^i \xi_i / 2) = \cosh(\xi/2)(1 \pm \sigma^i n_i \tanh(\xi/2))$ and the hyperbolic relations $\cosh(\xi/2) = \sqrt{(\cosh(\xi)+1)/2}$ and $\tanh(\xi/2) = \sinh(\xi)/(\cosh(\xi)+1)$ together with (13) and as usual $\sigma^i n_i = \frac{1}{|\boldsymbol{p}|}\begin{pmatrix} p_z & p_- \\ p_+ & -p_z \end{pmatrix}$, $p_\pm = p_x \pm ip_y$ yielding the deformed boost transformation

$$S = \sqrt{\frac{FE+m}{2m}} \begin{pmatrix} 1+\dfrac{Gp_z}{FE+m} & \dfrac{Gp_-}{FE+m} & 0 & 0 \\ \dfrac{Gp_+}{FE+m} & 1-\dfrac{Gp_z}{FE+m} & 0 & 0 \\ 0 & 0 & 1-\dfrac{Gp_z}{FE+m} & -\dfrac{Gp_-}{FE+m} \\ 0 & 0 & -\dfrac{Gp_+}{FE+m} & 1+\dfrac{Gp_z}{FE+m} \end{pmatrix} \tag{16}$$

The plane wave solutions of (14) in coordinate space are $\psi^+(x) = u(p)\exp(-ipx)$ and $\psi^-(x) = v(p)\exp(+ipx)$. In a rest frame we have $E = m$, $\boldsymbol{p} = 0$, $F = G = 1$ and $(\gamma^0 - 1)u(0) = 0$, $(\gamma^0 + 1)v(0) = 0$ which determines the solutions $u(0)$, $v(0)$ with the normalisation convention of [20] as

$$u_1(0) = \sqrt{m}\begin{pmatrix} 1 \\ 0 \\ 1 \\ 0 \end{pmatrix}, \ u_2(0) = \sqrt{m}\begin{pmatrix} 0 \\ 1 \\ 0 \\ 1 \end{pmatrix}, \ v_1(0) = \sqrt{m}\begin{pmatrix} 1 \\ 0 \\ -1 \\ 0 \end{pmatrix}, \ v_2(0) = \sqrt{m}\begin{pmatrix} 0 \\ 1 \\ 0 \\ -1 \end{pmatrix} \tag{17}$$

Then from (15),(16) and (17) we get the plane wave solutions of the DSR Dirac equation (14)

$$(u_1, u_2, v_1, v_2)(p) = \sqrt{\frac{FE+m}{2}} \left( \begin{array}{c} 1 + \frac{Gp_z}{FE+m} \\ \frac{Gp_+}{FE+m} \\ 1 - \frac{Gp_z}{FE+m} \\ -\frac{Gp_+}{FE+m} \end{array} \right), \left( \begin{array}{c} \frac{Gp_-}{FE+m} \\ 1 - \frac{Gp_z}{FE+m} \\ -\frac{Gp_-}{FE+m} \\ 1 + \frac{Gp_z}{FE+m} \end{array} \right), \left( \begin{array}{c} 1 + \frac{Gp_z}{FE+m} \\ \frac{Gp_+}{FE+m} \\ -1 + \frac{Gp_z}{FE+m} \\ \frac{Gp_+}{FE+m} \end{array} \right), \left( \begin{array}{c} \frac{Gp_-}{FE+m} \\ 1 - \frac{Gp_z}{FE+m} \\ \frac{Gp_-}{FE+m} \\ -1 - \frac{Gp_z}{FE+m} \end{array} \right)$$
(18)

Now from these solutions one can derive the various rules for the spinors $u_s$, $v_s$ (r,s=1,2) which we now write down, where $U(p_\mu) = (Fp_0, Gp_i)$:

$$\bar{u}_r(p)u_s(p) = 2m\delta_{rs}, \quad \bar{v}_r(p)v_s(p) = -2m\delta_{rs} \tag{19}$$
$$\bar{u}_r(p)v_s(p) = \bar{v}_r(p)u_s(p) = 0$$

$$\bar{u}_r(p)\gamma^\mu u_s(p) = \bar{v}_r(p)\gamma^\mu v_s(p) = 2U(p^\mu)\delta_{rs} \tag{20}$$
$$\bar{u}_r(p)\gamma^0 v_s(-p) = \bar{v}_r(p)\gamma^0 u_s(-p) = 0$$

$$\sum_s u_s(p)\bar{u}_s(p) = \gamma^\mu U(p_\mu) + m, \quad \sum_s v_s(p)\bar{v}_s(p) = \gamma^\mu U(p_\mu) - m \tag{21}$$

$$(\gamma^\mu U(p_\mu) - m)u(p) = 0, \quad (\gamma^\mu U(p_\mu) + m)v(p) = 0 \tag{22}$$

Many of the relations in this section could have been guessed from the special relativistic ones by transforming the unphysical variables $\pi$ to physical variables $p$ according to (2).

## 5. Space-Time formulation of a specific model

The deformed Dirac equation for a DSR theory in momentum space is (14). The coordinate space formulation is obtained via the replacement (6). Now we have to refer to a specific DSR theory and choose our model in [10], where $F = G = (1 - \ell^2 \boldsymbol{p}^2)^{-1/2}$ and $m$ automatically coincides with $m_0$. The deformed Dirac equation becomes

$$\frac{i\gamma^\mu \partial_\mu}{\sqrt{1 + \ell^2 \partial_j^2}} \psi(x) - m\psi(x) = 0 \tag{23}$$

where $\partial_j^2 = \nabla^2$. The corresponding Lagrangian is

$$L = \bar{\psi} \frac{i\gamma^\mu \partial_\mu}{\sqrt{1 + \ell^2 \partial_j^2}} \psi - m\bar{\psi}\psi = -(\partial_\mu \bar{\psi}) \frac{i\gamma^\mu}{\sqrt{1 + \ell^2 \partial_j^2}} \psi - m\bar{\psi}\psi \tag{24}$$

where the second expression is equivalent to the first one up to the 4-divergence $\partial_\mu(\overline{\psi}\frac{i\gamma^\mu}{\sqrt{1+\ell^2\partial_j^2}}\psi)$, which can be omitted. The first expression in (24) gives immediately the equation of motion (23) for $\psi$ since from (7) only the term $\frac{\partial L}{\partial \overline{\psi}}$ contributes. From the second expression in (24) and by Taylor expanding the square root one obtains via (7), where only the 1., 3., .. terms contribute: $-(\partial_\mu\overline{\psi})i\gamma^\mu - m\overline{\psi} + \frac{1}{2}\ell^2(\partial_j^2\partial_\mu\overline{\psi})i\gamma^\mu - \frac{3}{8}\ell^4(\partial_j^4\partial_\mu\overline{\psi})i\gamma^\mu ... = 0$. Collecting the Taylor expansion yields the equation of motion for $\overline{\psi}$

$$\frac{\partial_\mu}{\sqrt{1+\ell^2\partial_j^2}}\overline{\psi}(x)i\gamma^\mu + m\overline{\psi}(x) = 0 \tag{25}$$

Again from the second expression in (24) we can calculate the momentum density from (9b), where only the first term containing a time derivative and the last term contribute. One obtains then $T^{0\nu} = \partial^\nu\overline{\psi}\frac{\partial L}{\partial \partial_0\overline{\psi}} - L\eta^{0\nu} = -(\partial^\nu\overline{\psi})\frac{i\gamma^0}{\sqrt{1+\ell^2\partial_j^2}}\psi - L\eta^{0\nu}$. We are interested in the energy $H = \int d^3x\, T^{00} = \int d^3x[(\partial_i\overline{\psi})\frac{i\gamma^i}{\sqrt{1+\ell^2\partial_j^2}}\psi + m\overline{\psi}\psi] = \int d^3x[-\overline{\psi}\frac{i\gamma^i\partial_i}{\sqrt{1+\ell^2\partial_j^2}}\psi + m\overline{\psi}\psi]$, where we have omitted a 3-divergence in the last step after partial integration, see [21] p.139. Using the equation of motion (23) we get finally

$$H = \int d^3x\, T^{00} = \int d^3x\, \overline{\psi}\frac{i\gamma^0\partial^0}{\sqrt{1+\ell^2\partial_j^2}}\psi \tag{26a}$$

Similarly from $T^{0\nu}$ one receives for the 3-momentum after partial integration and neglecting a 3-divergence the following expression

$$P^i = \int d^3x\, T^{0i} = \int d^3x\, \overline{\psi}\frac{i\gamma^0\partial^i}{\sqrt{1+\ell^2\partial_j^2}}\psi \tag{26b}$$

The charge density $j^0$ can be calculated from the second form of the Lagrangian in (24) and the definition in (8), where only the first term contributes, as $j^0 = \delta\overline{\psi}\frac{\partial L}{\partial \partial_0\overline{\psi}}$ with $\delta\overline{\psi} = i\overline{\psi}$. After multiplying with the electric charge $e$ we get the conserved charge

$$Q = e\int d^3x\, j^0 = e\int d^3x\, \overline{\psi}\frac{\gamma^0}{\sqrt{1+\ell^2\partial_j^2}}\psi \tag{27}$$

In the present model eventually problematical time derivatives of order higher than two in a deformed Klein-Gordon equation are avoided. This is not possible in a theory with higher derivatives based on special relativity.

## 5. Second quantisation

In literature [16,19-22] there are several normalisation conventions in second quantisation, here we choose the one of [20] using the Weyl representation employed in section 4 and write down the plane wave expansion of the field operators accommodated to DSR.

$$\psi(x) = \sum_s \int \frac{d^3p}{(2\pi)^{3/2}} \frac{1}{\sqrt{2FE}} \left[ b_s(\boldsymbol{p}) u_s(\boldsymbol{p}) e^{-ipx} + d_s^+(\boldsymbol{p}) v_s(\boldsymbol{p}) e^{+ipx} \right]$$

$$\overline{\psi}(x) = \sum_s \int \frac{d^3p}{(2\pi)^{3/2}} \frac{1}{\sqrt{2FE}} \left[ b_s^+(\boldsymbol{p}) \overline{u}_s(\boldsymbol{p}) e^{+ipx} + d_s(\boldsymbol{p}) \overline{v}_s(\boldsymbol{p}) e^{-ipx} \right] \tag{28}$$

Note that the integration runs over a finite range due to $|\boldsymbol{p}| < 1/\ell$. The normalisation factor $1/\sqrt{2\varepsilon}$ of special relativity was transformed in $1/\sqrt{2FE}$, furthermore we use $(2\pi)^{3/2}$ instead of $(2\pi)^3$. The above normalisation of the field operators was chosen in order to give a Delta like normalisation for the anticommutators. As in [16] we postulate that the anticommutation relations between creation and annihilation operators must satisfy

$$\{b_r^+(\boldsymbol{p}'), b_s(\boldsymbol{p})\} = \{d_r^+(\boldsymbol{p}'), d_s(\boldsymbol{p})\} = \delta^{(3)}(\boldsymbol{p}' - \boldsymbol{p}) \delta_{rs} \tag{29}$$

with all other relations equal to zero. Now we calculate the equal time anticommutators between $\overline{\psi}$ and $\psi$ with (28), (29) and the deformed spin sums (21) with $U(p_\mu) = Fp_\mu$. In the final step we have to calculate $\int \frac{d^3p}{(2\pi)^3} \exp(i\boldsymbol{p}(\boldsymbol{x}' - \boldsymbol{x}))$, where the integration now runs over a finite volume since $|\boldsymbol{p}| < \kappa$. In one dimension we get a modified Delta function $\delta_\kappa(x) = \frac{1}{2\pi} \int_{-\kappa}^{\kappa} dp \exp(ipx) = \frac{\sin(\kappa \cdot x)}{\pi \cdot x}$, which is regular everywhere. $\delta_\kappa(x)$ has the following properties: (i) $\lim_{x \to 0} \delta_\kappa(x) = \kappa/\pi$ (which is very large), (ii) $\lim_{\kappa \to \infty} \delta_\kappa(x) = \delta(x)$ (it approximates the Dirac Delta function), (iii) $\int_{-\infty}^{\infty} \delta_\kappa(x) dx = 1$ (as for the Dirac Delta function).

However the Delta function in momentum space $\delta^{(3)}(\boldsymbol{p}) = \int \frac{d^3x}{(2\pi)^3} \exp(-i\boldsymbol{p}\boldsymbol{x})$ remains unmodified, since the integration runs over an infinite range. This yields in three dimensions as result for the anticommutator:

$$\{\overline{\psi}(\boldsymbol{x}',t), \psi(\boldsymbol{x},t)\} = \gamma^0 \delta_\kappa^{(3)}(\boldsymbol{x}' - \boldsymbol{x}) \tag{30}$$

The other anticommutators vanish automatically by virtue of (29).

Next we compute the energy given by (26a). We have to insert the plane wave expansion (28) and use $\frac{i\gamma^0 \partial^0}{\sqrt{1 + \ell^2 \partial_j^2}} e^{\pm ipx} = \mp \gamma^0 FE e^{\pm ipx}$. With the deformed spinor relations (20) and the unmodified $\delta^{(3)}(\boldsymbol{p})$ we get for the energy $H = \sum_s \int d^3p \ FE[b_s^+(\boldsymbol{p}) b_s(\boldsymbol{p}) - d_s(\boldsymbol{p}) d_s^+(\boldsymbol{p})]$ or with (29) the final result for the energy in second quantisation:

$$H = \sum_s \int d^3p \ FE[b_s^+(\boldsymbol{p})b_s(\boldsymbol{p}) + d_s^+(\boldsymbol{p})d_s(\boldsymbol{p})] - \delta^{(3)}(0)\sum_s \int d^3p \ FE \qquad (31)$$

Neglecting at first the second term with the vacuum contribution, the expression for the energy $H$ looks like the one in special relativity (SR) where the energy $E$ is replaced by $FE$ and therefore enhanced, however the integration runs over a finite volume in momentum space. For the factor $\delta^{(3)}(0)$ in the second term see the discussion in [22]. Consider the integral $E_0 = -\int d^3p \ FE$, which in special relativity with $F=1$ constitutes the divergent vacuum energy. From the dispersion relation (1) we have $FE = \sqrt{G^2 \boldsymbol{p}^2 + m^2} = \sqrt{\dfrac{\boldsymbol{p}^2}{1-\ell^2\boldsymbol{p}^2} + m^2}$ and if we approximately omit $m$, which is very small compared to the other term, the integral becomes: (the exact value involves elliptic functions)

$$E_0 = -\int d^3p \ FE \approx -\int d^3p \sqrt{\dfrac{\boldsymbol{p}^2}{1-\ell^2\boldsymbol{p}^2}} = -4\pi \int_0^{1/\ell} d\rho \dfrac{\rho^3}{\sqrt{1-\ell^2\rho^2}} = -\dfrac{8\pi}{3\ell^4} \qquad (32)$$

So in contrary to special relativity we get in our model a finite vacuum energy finally produced by the deformed dispersion relation with $F = G = (1-\ell^2\boldsymbol{p}^2)^{-1/2}$.

In analogy to $H$ the three momentum $P^i$, where the vacuum term does not contribute on symmetry grounds, becomes from (26b)

$$P^i = \sum_s \int d^3p \ Fp^i[b_s^+(\boldsymbol{p})b_s(\boldsymbol{p}) + d_s^+(\boldsymbol{p})d_s(\boldsymbol{p})] \qquad (33)$$

Finally from (27) the conserved charge $Q$ becomes

$$Q = e\sum_s \int d^3p \ F[b_s^+(\boldsymbol{p})b_s(\boldsymbol{p}) - d_s^+(\boldsymbol{p})d_s(\boldsymbol{p})] + e\delta^{(3)}(0)\sum_s \int d^3p \ F \qquad (34)$$

Besides the vacuum term the charge is enlarged or antiscreened compared to the SR value, a result also obtained in [23] from the construction of deformed Dirac spinors in DSR models. For the vacuum contribution to the charge again we get a finite value due to

$$Q_0 = e\int d^3p \ F = e\int d^3p \dfrac{1}{\sqrt{1-\ell^2\boldsymbol{p}^2}} = e4\pi \int_0^{1/\ell} d\rho \dfrac{\rho^2}{\sqrt{1-\ell^2\rho^2}} = \dfrac{e\pi^2}{\ell^3} \qquad (35)$$

Note that the particle antiparticle interpretation of the creation and annihilation operators as well as the spin statistics connection remains valid. It is quite exiting that the present approach yields finite expressions for the vacuum energy and charge and thus eliminates one of the divergences of standard quantum field theory, even if this is the least worse one. From the view of DSR theories the reason for the infinite vacuum contributions resides in special relativity with its dispersion relation, where energy and momentum can have infinitely high values.

## 6. Summary

In summary we have considered a field theoretical formulation of DSR theories based on a commutative space-time, which could at least serve as a first approximation in the case of a noncommutative space-time. In general the corresponding field theories contain derivatives of arbitrary order and one has to apply the higher derivatives field theory formulation. Based on the deformed Dirac equation and plane wave solutions in momentum space we obtained for the DSR model in [10] expressions for momentum and charge in a quantisation procedure. They show up an enhanced momentum and charge and finite values for the vacuum contributions, contrary to special relativity with infinite vacuum values. The integrals in (32) and (35) bear a formal similarity to an old work of Born [24] (see also [25]), who introduced an ad hoc cutoff in momentum space, however the interpretation is quite different.

There are several possible extensions to the present work. One should investigate interactions in the commutative framework. Of equal interest is a formulation in noncommutative space-time. A central question is of course: can the extremely unsatisfactory divergences of standard quantum field theory based on special relativity be avoided in DSR theories?